\documentclass[12pt]{article}
\usepackage{amssymb,stmaryrd,amsmath,amsfonts,rotating}%amsfonts,amsthm 
\usepackage[noadjust]{cite}
\usepackage{color}%temporary packages
\usepackage[vflt]{floatflt}
\usepackage{epic}

\definecolor{TODO}{rgb}{0.6,0.6,0.6} % TO DO!!!!

\definecolor{TOCHECK}{rgb}{0.8,0.8,0.8} % TO CKECK!!!!

\newtheorem{theorem}{Theorem}
\newcommand{\btheo}{\begin{theorem}}
\newcommand{\etheo}{\end{theorem}}
\newtheorem{proof}{Proof}
\newcommand{\bproof}{\begin{proof}}
\newcommand{\eproof}{\end{proof}}
\newtheorem{definition}{Definition}
\newcommand{\bdefi}{\begin{definition}}
\newcommand{\edefi}{\end{definition}}
\newtheorem{fact}{Fact}
\newcommand{\bprop}{\begin{fact}}
\newcommand{\eprop}{\end{fact}}
\newtheorem{corollary}{Corollary}
\newcommand{\bcor}{\begin{corollary}}
\newcommand{\ecor}{\end{corollary}}
\newtheorem{example}{Example}
\newcommand{\bex}{\begin{example}}
\newcommand{\eex}{\end{example}}
\newtheorem{lemma}{Lemma}
\newcommand{\blemma}{\begin{lemma}}
\newcommand{\elemma}{\end{lemma}}
\newtheorem{remark}{Remark}
\newcommand{\bremark}{\begin{remark}}
\newcommand{\eremark}{\end{remark}}
\newcommand{\naturals}{\ensuremath{\mathbb{N}}}
\newcommand{\defas}{\ensuremath{\stackrel{{\vartriangle}}{=}}} 
\def\0{{\tt 0}} % Ex.: BPSK modulation => 0 is encoded into +1
\def\1{{\tt 1}} % Ex.: BPSK modulation => 1 is encoded into -1
\def\?{{\tt *}} % erasure symbol
\newcommand{\ledge}{\ensuremath{\lambda}} % ledge (edge perspective) 
\newcommand{\redge}{\ensuremath{\rho}} % redge (edge perspective
\newcommand{\cp}{\epsilon} % channel parameter 
\newcommand{\MAP}{\ensuremath{\text{MAP}}} % Maximum A Posteriori
\newcommand{\BP}{\ensuremath{\text{BP}}} % Belief Propagation 
\newcommand{\Shsmall}{\ensuremath{\text{\tiny Sh}}} % Shannon
\newcommand{\dens}[1]{\mathsf{\MakeLowercase{#1}}}
\newcommand{\Ldens}[1]{\dens{#1}}

\newcommand{\BEC}{\ensuremath{\text{BEC}}}

\newcommand{\BSC}{\ensuremath{\text{BSC}}}

\newcommand{\gexitkl}[2]{l^{#1}(#2)}  % GEXIT kernel in L domain
\def\0t{{\tt 0}}
\def\1t{{\tt 1}}

\newcommand{\entropy}{H} % entropy operator 
\newcommand{\gentropy}{G} % entropy operator
\newcommand{\conv}{\star}
\newcommand{\ent}{\ensuremath{{\tt{h}}}}

\newcommand{\lognat}{\log}

\begin{document}
\renewcommand{\textfraction}{0}

\title{Why We Can Not Surpass Capacity: The Matching Condition 
%\footnote{The work of A.~Montanari was partially supported by the European Union under 
%the project EVERGROW.}
%
%This research is
%supported in part by NSF Grant No.~NCR-9627610}
} \author{\normalsize Cyril M\'easson \\ 
\small EPFL \\[-5pt] 
\small CH-1015 Lausanne \\[-5pt]
\small cyril.measson@epfl.ch, 
\and
\normalsize Andrea Montanari \\
%supported in part by NSF Grant No.~NCR-9627610} \\
\small LPT, ENS \\[-5pt]
\small F-75231 Paris \\[-5pt]
\small montanar@lpt.ens.fr
\and
\normalsize R\"udiger Urbanke \\
\small EPFL \\[-5pt] 
\small CH-1015 Lausanne \\[-5pt]
\small ruediger.urbanke@epfl.ch 
}
\date{}
\maketitle
\thispagestyle{empty}
\begin{abstract}
We show that iterative coding systems can not surpass capacity {\em using only quantities
which naturally appear in density evolution.} Although the result in itself is 
trivial, the method which we apply shows that in order to achieve capacity
the various components in an iterative coding system have to be perfectly matched. 
This generalizes the perfect matching condition which was previously known for the
case of transmission over the binary erasure channel to the general class of binary-input memoryless
output-symmetric channels. Potential applications of this perfect matching condition
are the construction of capacity-achieving degree distributions and the determination
of the number required iterations as a function of
the multiplicative gap to capacity.
\end{abstract}
\normalsize

Assume we are transmitting over a binary-input memoryless output-symmetric (BMS) channel
using sparse graph codes and an iterative decoder. Why can we not surpass capacity with
such a set-up? The trivial answer is of course given by the converse to the channel
coding theorem. In this paper we give an alternative proof of this fact which uses
only quantities which are naturally tied to the setup of iterative coding.
We show that in order to achieve capacity
the various components in an iterative coding system have to be perfectly matched.
This generalizes the perfect matching condition which was previously known for the
case of transmission over the binary erasure channel (BEC) to the general class of BMS 
channels.

The first bound in which capacity was derived explicitly from density evolution
was given by Shokrollahi and Oswald \cite{Sho00,OsS01} for the case of
transmission over the BEC. For the same channel, a very pleasing geometric bound using
ten Brink's EXIT charts (\cite{teB99a,teB99b,teB00,teB01}) was later given
by Ashikhmin, Kramer and ten Brink using the Area Theorem \cite{AKtB02a,AKTB02,AKtB04}.
See also \cite{HHJF01,HuH02b,MeU02jccc,MeU03a} for 
related work and the extension to parallel concatenation. 

For general BMS channels, 
this geometric interpretation is unfortunately no longer valid since the
Area Theorem is no longer fulfilled by the
component-wise  EXIT curves. Motivated by the
pleasing geometric statement observed for the 
BEC, a similar chart, 
called MSE chart was constructed by Bhattad and Narayanan \cite{BaN04}.
Assuming that the input densities to the component codes are Gaussian, this
chart again fulfills the Area Theorem. The introduction of this
function was motivated by the elegant 
relationship between mutual information and signal-to-noise 
observed by Guo, Shamai and Verdu \cite{GSV04,GSV05}.
In order to apply the MSE chart in the context of iterative coding
the authors proposed to approximate the intermediate densities which
appear in density evolution by ``equivalent'' Gaussian densities.
This was an important first step in generalizing the matching condition
to the whole class of BMS channels.  

In the following we show how to overcome the need for making the
Gaussian approximation by using generalized EXIT (GEXIT) functions \cite{MMRU04}.
The bound which we derive and its geometric interpretation is quite
similar to the one given for the BEC: We represent the ``actions''
performed by each component code by their respective GEXIT functions.
By construction the area under these GEXIT functions is related to the rate
of the corresponding codes. Further, we show that if we are transmitting below 
the threshold of iterative coding then these two curves do not overlap.
Using then an argument identical to the one introduced for the BEC it follows
that in order to achieve capacity the two individual component curves have to
be perfectly matched.

There are two obvious potential applications of the perfect matching condition.
First, assuming a perfect matching of the component curves and working backwards,
one might be able to exhibit capacity-achieving degree distribution pairs.
Secondly, the geometric picture given by the bound seems to provide the natural 
setting to prove that the number of iterations of iterative codings systems
scales at least like $\Theta(1/\delta)$, where $\delta$ is the multiplicative 
gap to capacity, another long standing conjecture of iterative coding.

\section{EXIT Charts and the Matching Condition for BEC}
\begin{figure}[hbt]
\centering
\setlength{\unitlength}{1.0bp}%
\begin{picture}(110,110)
\put(0,0){\includegraphics[scale=1.0]{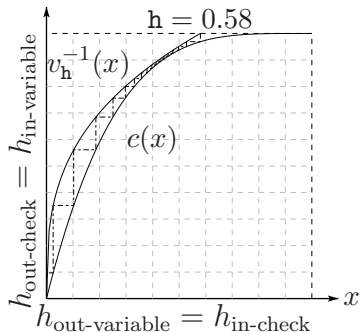}}
%\put(100,-3){\makebox(0,0)[t]{\small $1$}}
%\put(-5,102){\makebox(0,0){\small $1$}}
\put(112, 0){\makebox(0,0)[l]{\small $x$}}
%\put(18, 93){\makebox(0,0){\small $\ledge^{-1}(x/\ent)$}}
\put(40, 60){\makebox(0,0){\small $c(x)$}}
\put(16, 89){\makebox(0,0){\small $v^{-1}_{\ent}(x)$}}
\put(58, 102){\makebox(0,0)[b]{{\small $\ent=0.58$}}}
\put(100, -2){\makebox(0,0)[tr]{$h_{\text{out-variable}}=h_{\text{in-check}}$}}
\put(-4,100){\makebox(0,0)[tr]{\rotatebox{90}{$h_{\text{out-check}}=h_{\text{in-variable}}$}}}
\end{picture}
\caption{\label{fig:becmatching} The EXIT chart method for
the degree distribution $(\ledge(x)=x^3, \redge(x)=x^4)$ and
transmission over the $\BEC(\ent = 0.58)$.}
\end{figure}

To start, let us review the case of transmission over the $\BEC(\ent)$
using a degree distribution pair $(\ledge, \redge)$.
In this case density evolution is equivalent to the EXIT chart
approach and the condition for successful decoding under \BP\ reads
\begin{align*}
c(x) \defas 1-\redge(1-x) \leq \ledge^{-1}(x/\ent) \defas v^{-1}_{\ent}(x).
\end{align*}
This is shown in Fig.~\ref{fig:becmatching} for the degree distribution pair
$(\ledge(x)=x^3, \redge(x)=x^4)$. 
The area under the curve $c(x)$ equals $1-\int \!\redge$ and the 
area to the left of the curve $v^{-1}_{\ent}(x)$ is equal to 
$\ent \int \!\ledge$. By the previous remarks, a necessary condition
for successful \BP\ decoding
is that these two areas do not overlap.
Since the total area equals $1$ we get the necessary condition
$\ent \int \ledge+1-\int \redge\leq 1$. Rearranging terms, this 
is equivalent to the condition
\begin{align*}
1-C_{\Shsmall} = \ent \leq \frac{\int \redge}{\int \ledge}= 1 - r(\ledge, \redge).
\end{align*}
In words, the rate $r(\ledge, \redge)$ of any LDPC ensemble which, for
increasing block lengths, allows successful 
decoding over the $\BEC(\ent)$, can not surpass the Shannon limit
$1-\ent$. 
As pointed out in the introduction, an argument very similar to the above was introduced
by Shokrollahi and Oswald \cite{Sho00,OsS01} (albeit not using the language and geometric
interpretation of EXIT functions and applying a slightly different range of integration).
It was the first bound on the performance of iterative systems in which the Shannon capacity
appeared explicitly using only quantities of density evolution.
A substantially more general version of this bound can be found in \cite{AKtB02a,AKTB02,AKtB04}
(see also Forney \cite{For05}).

Although the final result (namely that transmission above capacity
is not possible) is trivial, the method of proof is well worth the effort
since it shows how capacity enters in the calculation of the performance 
of iterative coding systems. By turning this bound around, we 
can find conditions under which iterative systems achieve capacity:
In particular it shows that the two component-wise
EXIT curves have to be matched perfectly. Indeed, all currently known
capacity achieving degree-distributions for the BEC can be derived 
by starting with this perfect matching condition and working backwards.

\section{GEXIT Charts and the Matching Condition for BMS Channels}
Let us now derive the equivalent result for general BMS channels.
As a first ingredient we show how to interpolate the sequence of densities
which we get from density evolution so as to form a complete family of
densities.
%GEXIT functions have been introduced in \cite{MMRU04} and fulfilled 
%the area theorem by definition over any BMS channel.  
%In previous works \cite{MMRU04,MMRU05} we have so far mainly 
%used GEXIT functions to derive an upper-bound on
%the \MAP\ threshold of iterative coding systems by substituting
%the \BP\ for the \MAP\ GEXIT function. Of course, by referring to Shannon's 
%channel coding theorem, we know that this threshold can not surpass
%the Shannon threshold. Let us now give an alternative proof of this
%fact which uses only basic facts of GEXIT functions and does not
%refer to Shannon's coding theorem.
%
%While our previous work uses GEXIT functions of the overall coding systems 
%(in the asymptotic limit when the blocklength tends to infinity), we 
%consider now and in the sequel GEXIT functions of individual component codes.
%
%
%Let us now show that, by using component-wise GEXIT functions, the perfect
%matching condition holds in the general case. This might in the future serve 
%as a starting point to find capacity-achieving degree distributions for
%general BMS channels. We need some preliminary definitions and lemmas.

\bdefi[Interpolating Channel Families]
\label{def:interpolation}
Consider a degree distribution pair $(\ledge, \redge)$
and transmission over the BMS channel characterized by its
$L$-density $\Ldens{c}$. Let $\Ldens{a}_{-1}=\Delta_0$
and $\Ldens{a}_0=\Ldens{c}$ and set $\Ldens{a}_{\alpha}$,
$\alpha \in [-1, 0]$, to 
$\Ldens{a}_{\alpha}=-\alpha \Ldens{a}_{-1} + (1+\alpha) \Ldens{a}_0$.
The {\em interpolating density evolution families} 
$\{\Ldens{a}_{\alpha}\}_{\alpha=-1}^{\infty}$
and $\{\Ldens{b}_{\alpha}\}_{\alpha=0}^{\infty}$ are then defined as follows:
\begin{align*}
\Ldens{b}_{\alpha} & = \sum_{i} \redge_i \Ldens{a}_{\alpha-1}^{\boxast (i-1)},
\;\;\;\;\; \alpha \geq 0,\\
\Ldens{a}_{\alpha} & =  
\sum_{i} \ledge_i \Ldens{c} \conv \Ldens{b}_{\alpha}^{\conv (i-1)},
\;\;\;\;\;\alpha \geq 0,
\end{align*}
where $\conv$ denotes the standard convolution of densities and
$\Ldens{a} \boxast \Ldens{b}$ denotes the density at the output of
a check node, assuming that the input densities are $\Ldens{a}$ and $\Ldens{b}$,
respectively.
\edefi
Discussion: First note that $\Ldens{a}_{\ell}$ ($\Ldens{b}_{\ell}$), 
$\ell \in \naturals$,
represents the sequence of $L$-densities of density evolution
emitted by the variable (check) nodes in the $\ell$-th iteration.
By starting density evolution not only with $\Ldens{a}_{0}=\Ldens{c}$
but with all possible convex combinations of $\Delta_0$ and
$\Ldens{c}$, this discrete sequence of densities is completed to
form a continuous family of densities ordered by physical degradation.
The fact that the densities are ordered by physical degradation
can be seen as follows: note that the computation tree for $\Ldens{a}_{\alpha}$ 
can be constructed by taking  
the standard computation tree of $\Ldens{a}_{\lceil \alpha \rceil}$ 
and independently erasing the observation associated to each variable leaf node with probability
$\lceil \alpha \rceil-\alpha$. It follows that we can convert the computation tree of 
$\Ldens{a}_{\alpha}$ to that of $\Ldens{a}_{\alpha-1}$ by erasing all
observations at the leaf nodes and by independently erasing
each observation in the second (from the bottom) row of variable nodes
with probability $\lceil \alpha \rceil-\alpha$.
The same statement is true for $\Ldens{b}_{\alpha}$.
If $\lim_{\ell \rightarrow \infty} \entropy(\Ldens{a}_{\ell})=0$, i.e., 
if \BP\
decoding is successful in the limit of large blocklengths, then
the families are both complete.

\begin{example}[Density Evolution and Interpolation]
Consider transmission over the $\BSC(0.07)$ using a
$(3, 6)$-regular ensemble. Fig.~\ref{fig:debsc} depicts
the density evolution process for this case.
\begin{figure}[htp]
\setlength{\unitlength}{0.5bp}%
\begin{center}
\begin{picture}(740,170)
\put(0,0)
{
\put(0,0){\includegraphics[scale=0.5]{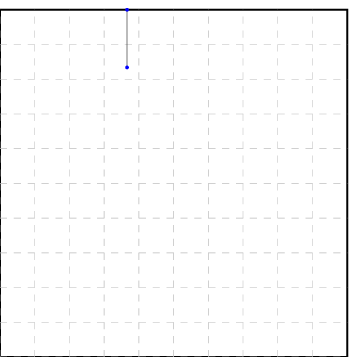}}
\put(120,50){\includegraphics[scale=0.5]{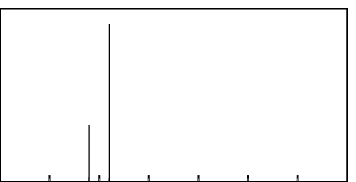}}
\put(0,120){\includegraphics[scale=0.5]{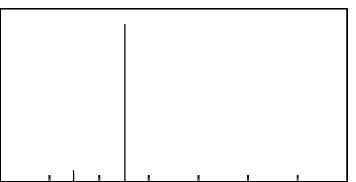}}
\put(50, 110){\makebox(0,0)[c]{\tiny $a_{0}$}}
\put(170, 40){\makebox(0,0)[c]{\tiny $b_{1}$}}
\put(50, -2){\makebox(0,0)[t]{\tiny $\entropy(\Ldens{a})$}}
\put(-2, 50){\makebox(0,0)[r]{\tiny \rotatebox{90}{$\entropy(\Ldens{b})$}}}
}
\put(260,0)
{
\put(0,0){\includegraphics[scale=0.5]{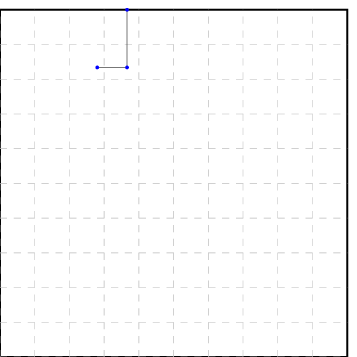}}
\put(120,50){\includegraphics[scale=0.5]{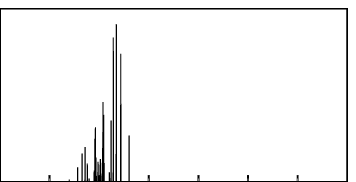}}
\put(0,120){\includegraphics[scale=0.5]{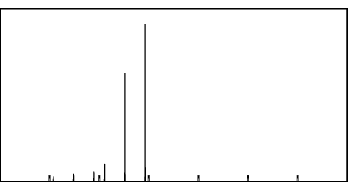}}
\put(50, 110){\makebox(0,0)[c]{\tiny $a_{1}$}}
\put(170, 40){\makebox(0,0)[c]{\tiny $b_{2}$}}
\put(50, -2){\makebox(0,0)[t]{\tiny $\entropy(\Ldens{a})$}}
\put(-2, 50){\makebox(0,0)[r]{\tiny \rotatebox{90}{$\entropy(\Ldens{b})$}}}
}
\put(500,0)
{
\put(0,0){\includegraphics[scale=0.5]{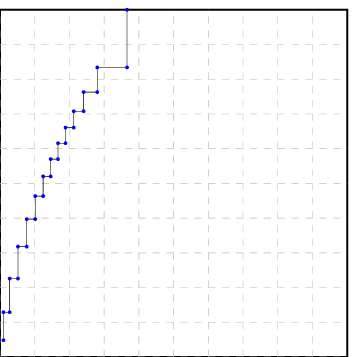}}
\put(120,50){\includegraphics[scale=0.5]{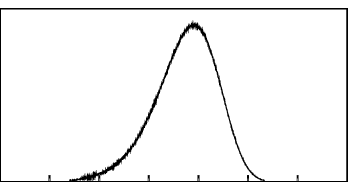}}
\put(0,120){\includegraphics[scale=0.5]{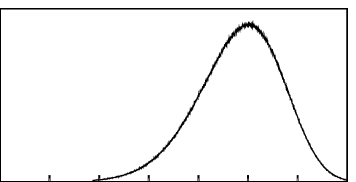}}
\put(50, 110){\makebox(0,0)[c]{\tiny $a_{12}$}}
\put(170, 40){\makebox(0,0)[c]{\tiny $b_{13}$}}
\put(50, -2){\makebox(0,0)[t]{\tiny $\entropy(\Ldens{a})$}}
\put(-2, 50){\makebox(0,0)[r]{\tiny \rotatebox{90}{$\entropy(\Ldens{b})$}}}
}
\end{picture}
\end{center}
\caption{\label{fig:debsc} Density evolution for $(3, 6)$-regular ensemble over $\BSC(0.07)$.}
\end{figure}
This process gives rise to the sequences of densities $\{\Ldens{a}_{\ell}\}_{\ell =0}^{\infty}$,
and $\{ \Ldens{b}_{\ell}\}_{\ell=1}^{\infty}$. Fig.~\ref{fig:interpolation} shows
the interpolation of these sequences for the choices $\alpha=1.0, 0.95, 0.9$ and $0.8$
and the complete such family.
\begin{figure}[htp]
\setlength{\unitlength}{0.6bp}%
\begin{center}
\begin{picture}(650,110)
\put(0,0){\includegraphics[scale=0.6]{de25}}
\put(50, 102){\makebox(0,0)[b]{\tiny $\alpha=1.0$}}
\put(50, -2){\makebox(0,0)[t]{\tiny $\entropy(\Ldens{a})$}}
\put(-2, 50){\makebox(0,0)[r]{\tiny \rotatebox{90}{$\entropy(\Ldens{b})$}}}
\put(130,0){\includegraphics[scale=0.6]{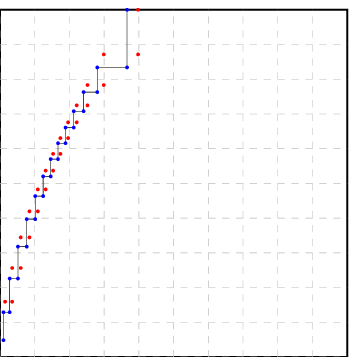}}
\put(180, 102){\makebox(0,0)[b]{\tiny $\alpha=0.95$}}
\put(180, -2){\makebox(0,0)[t]{\tiny $\entropy(\Ldens{a})$}}
\put(108, 50){\makebox(0,0)[r]{\tiny \rotatebox{90}{$\entropy(\Ldens{b})$}}}
\put(260,0){\includegraphics[scale=0.6]{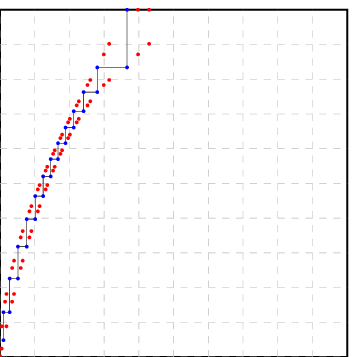}}
\put(310, 102){\makebox(0,0)[b]{\tiny $\alpha=0.9$}}
\put(310, -2){\makebox(0,0)[t]{\tiny $\entropy(\Ldens{a})$}}
\put(258, 50){\makebox(0,0)[r]{\tiny \rotatebox{90}{$\entropy(\Ldens{b})$}}}
\put(390,0){\includegraphics[scale=0.6]{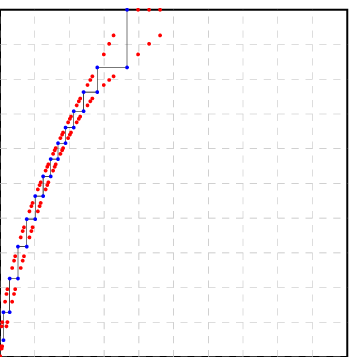}}
\put(440, 102){\makebox(0,0)[b]{\tiny $\alpha=0.8$}}
\put(440, -2){\makebox(0,0)[t]{\tiny $\entropy(\Ldens{a})$}}
\put(388, 50){\makebox(0,0)[r]{\tiny \rotatebox{90}{$\entropy(\Ldens{b})$}}}
\put(520,0){\includegraphics[scale=0.6]{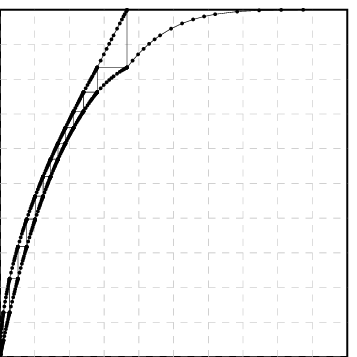}}
\put(570, -2){\makebox(0,0)[t]{\tiny $\entropy(\Ldens{a})$}}
\put(518, 50){\makebox(0,0)[r]{\tiny \rotatebox{90}{$\entropy(\Ldens{b})$}}}
\end{picture}
\end{center}
\caption{\label{fig:interpolation} Interpolation of densities.}
\end{figure}
\end{example}

As a second ingredient we recall from \cite{MMRU04} the definition of GEXIT functions. 
These GEXIT functions fulfill the Area Theorem for the case of general
BMS channels.
Up to date, GEXIT functions have been mainly
used to derive upper bounds on the \MAP\ threshold of iterative
coding systems, see e.g., \cite{MMRU04,MMRU05}. Here we will apply them
to the components of LDPC ensembles.

\bdefi[The GEXIT Functional]
Given two families of $L$-densities
$\{\Ldens{c}_\cp\}$ and $\{\Ldens{a}_\cp\}$ parameterized by $\epsilon$ define
the GEXIT functional $\gentropy(\Ldens{c}_{\cp}, \Ldens{a}_{\cp})$ by
\begin{align*}
\gentropy(\Ldens{c}_{\cp}, \Ldens{a}_{\cp}) & =
\int_{-\infty}^{\infty} \Ldens{a}_{\cp}(z) \gexitkl {\Ldens{c}_{\cp}} z \text{d}z,
\end{align*}
where
\begin{align*}
\gexitkl {\Ldens{c}_{\cp}} z
& =
\frac{\int_{-\infty}^{\infty} \frac{\text{d} \Ldens{c}_{\cp}(w)}{\text{d} \cp} 
\lognat(1+e^{-z-w}) \text{d}w}{
\int_{-\infty}^{\infty} \frac{\text{d} \Ldens{c}_{\cp}(w)}{\text{d} \cp}
\lognat(1+e^{-w}) \text{d}w}.
\end{align*}
Note that the kernel is normalized not with respect to $d \epsilon$ but
with respect to $d \ent$, i.e., with respect to changes in the entropy.
The families are required to be smooth in the sense that 
$\{\entropy(\Ldens{c}_{\cp}), \gentropy(\Ldens{c}_{\cp}, \Ldens{a}_{\cp})\}$
forms a piecewise continuous curve.
\edefi
\blemma[GEXIT and Dual GEXIT Function]
\label{lem:dualgexit}
Consider a binary code $C$ and transmission over a complete
family of BMS channels characterized by their family of $L$-densities
$\{\Ldens{c}_\cp\}$. Let $\{\Ldens{a}_{\cp}\}$ denote
the corresponding family of (average) extrinsic \MAP\ densities.
Then the standard GEXIT curve is given in parametric form by
$\{\entropy(\Ldens{c}_{\cp}), \gentropy(\Ldens{c}_{\cp}, \Ldens{a}_{\cp})\}$.
The {\em dual}
GEXIT curve is defined by 
$\{\gentropy(\Ldens{a}_{\cp}, \Ldens{c}_{\cp}), \entropy(\Ldens{a}_{\cp})\}$. 
Both, standard and dual GEXIT curve have an area equal to 
$r(C)$, the rate of the code.
\elemma
Discussion:
Note that both curves are ``comparable'' in that they first component measures
the channel $\Ldens{c}$ and the second argument measure the \MAP\ density
$\Ldens{a}$. The difference between the two 
lies in the choice of measure which is applied to each component.

\begin{proof}
Consider the entropy
$\entropy(\Ldens{c}_{\cp} \conv \Ldens{a}_{\cp})$. We have
\begin{align*}
\entropy(\Ldens{c}_{\cp} \conv \Ldens{a}_{\cp}) & =
\int_{-\infty}^{\infty} \Bigl(\int_{-\infty}^{\infty}
\Ldens{c}_{\cp}(w) \Ldens{a}_{\cp}(v-w) \text{d}w \Bigr) \log(1+e^{-v}) \text{d}v \\
& = \int_{-\infty}^{\infty} \int_{-\infty}^{\infty}
\Ldens{c}_{\cp}(w) \Ldens{a}_{\cp}(z) \log(1+e^{-w-z}) \text{d}w \text{d}z
\end{align*}
Consider now $\frac{\text{d} \entropy(\Ldens{c}_{\cp} \conv \Ldens{a}_{\cp})}{\text{d} \cp}$.
Using the previous representation we get
\begin{align*}
\frac{\text{d} \entropy(\Ldens{c}_{\cp} \conv \Ldens{a}_{\cp})}{\text{d} \cp} & =
\int_{-\infty}^{\infty} \int_{-\infty}^{\infty}
\frac{\text{d}\Ldens{c}_{\cp}(w)}{\text{d} \cp} \Ldens{a}_{\cp}(z) \log(1+e^{-w-z}) \text{d}w \text{d}z + \\
& \phantom{=} \int_{-\infty}^{\infty} \int_{-\infty}^{\infty}
\Ldens{c}_{\cp}(w) \frac{\text{d} \Ldens{a}_{\cp}(z)}{\text{d} \cp} \log(1+e^{-w-z}) \text{d}w \text{d}z.
\end{align*}
The first expression can be identified with the standard GEXIT curve 
except that it is parameterized by a generic parameter $\cp$.
The second expression is essentially the same, but the roles of
the two densities are exchanged.

Integrate now this relationship over the whole range of $\cp$ and
assume that this range goes from ``perfect'' (channel) to ``useless''. 
The integral on the left clearly equals 1. To perform the integrals
over the right reparameterize the first expression with respect to 
$\ent \defas \int_{\infty}^{\infty} \Ldens{c}_{\cp}(w) \lognat(1+e^{-w}) \text{d} w$
so that it becomes the standard GEXIT curve given by 
$\{\entropy(\Ldens{c}_{\cp}), \gentropy(\Ldens{c}_{\cp}, \Ldens{a}_{\cp})\}$.
In the same manner reparameterize the second expression by
$\ent \defas \int_{\infty}^{\infty} \Ldens{a}_{\cp}(w) \lognat(1+e^{-w}) \text{d} w$
so that it becomes the curve given by
$\{\entropy(\Ldens{a}_{\cp}), \gentropy(\Ldens{a}_{\cp}, \Ldens{c}_{\cp})\}$.
Since the sum of the two areas equals one and the area under the 
standard GEXIT curve equals $r(C)$, it follows that the area under
the second curve equals $1-r(C)$. Finally, note that if we consider the inverse
of the second curve by exchanging the two coordinates, i.e., if we consider the
curve 
$\{\gentropy(\Ldens{a}_{\cp}, \Ldens{c}_{\cp}), \entropy(\Ldens{a}_{\cp})\}$,
then the area under this curve is equal to $1-(1-r(C))=r(C)$, as claimed. 
\end{proof}
\begin{example}[EXIT Versus GEXIT]
Fig.~\ref{fig:exitversusgexit} compares the EXIT function to the
GEXIT function for the $[3,1,3]$ repetition code and the $[6,5,2]$ single parity-check
code when transmission takes place over the \BSC. As we can see, the two curves
are similar but distinct. In particular note that the areas
under the GEXIT curves are equal to the rate of the codes but that this is not
true for the EXIT functions.
\begin{figure}[htp]
\setlength{\unitlength}{1.0bp}%
\begin{center}
\begin{picture}(300,120)
\put(0,0)
{
\put(0,0){\includegraphics[scale=1.0]{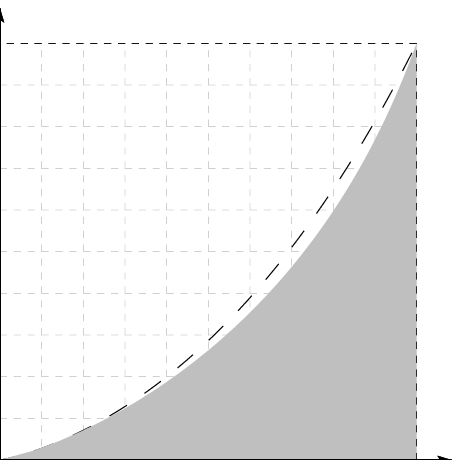}}
\put(90,30){\makebox(0,0)[c]{\small $\frac13$}}
\put(60, -2){\makebox(0,0)[t]{\small $\entropy(\Ldens{c}_{\ent})$}}
\put(-2, 60){\makebox(0,0)[r]{\small \rotatebox{90}{$\entropy(\Ldens{a}_{\ent}), \gentropy(\Ldens{c}_{\ent}, \Ldens{a}_{\ent})$}}}
\put(-2, -2){\makebox(0,0)[rt]{\small $0$}}
\put(120,-2){\makebox(0,0)[t]{\small $1$}}
\put(-2,120){\makebox(0,0)[r]{\small $1$}}
\put(60,80){\makebox(0,0)[b]{\small $[3, 1, 3]$}}
}
\put(180,0)
{
\put(0,0){\includegraphics[scale=1.0]{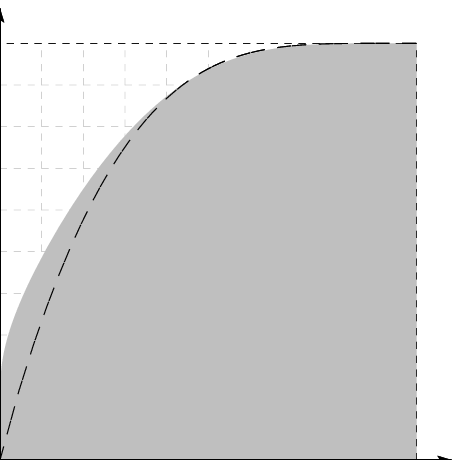}}
\put(90,30){\makebox(0,0)[c]{\small $\frac56$}}
\put(60, -2){\makebox(0,0)[t]{\small $\entropy(\Ldens{c}_{\ent})$}}
\put(-2, 60){\makebox(0,0)[r]{\small \rotatebox{90}{$\entropy(\Ldens{a}_{\ent}), \gentropy(\Ldens{c}_{\ent}, \Ldens{a}_{\ent})$}}}
\put(-2, -2){\makebox(0,0)[rt]{\small $0$}}
\put(120,-2){\makebox(0,0)[t]{\small $1$}}
\put(-2,120){\makebox(0,0)[r]{\small $1$}}
\put(60,40){\makebox(0,0)[b]{\small $[6, 5, 2]$}}
}
\end{picture}
\end{center}
\caption{\label{fig:exitversusgexit} A comparison of the EXIT with the GEXIT function for the
$[3,1,3]$ and the $[6, 5, 2]$ code.} 
\end{figure}
\end{example}
\begin{example}[GEXIT Versus Dual GEXIT]
Fig.~\ref{fig:gexitanddualgexit} shows the standard
GEXIT function and the dual GEXIT function for the $[5, 4, 2]$ code
and transmission over the $\BSC$. Although the two curves have quite
distinct shapes, the area under the two curves is the same.
\begin{figure}[htp]
\setlength{\unitlength}{1.0bp}%
\begin{center}
\begin{picture}(400,120)
\put(0,0)
{
\put(0,0){\includegraphics[scale=1.0]{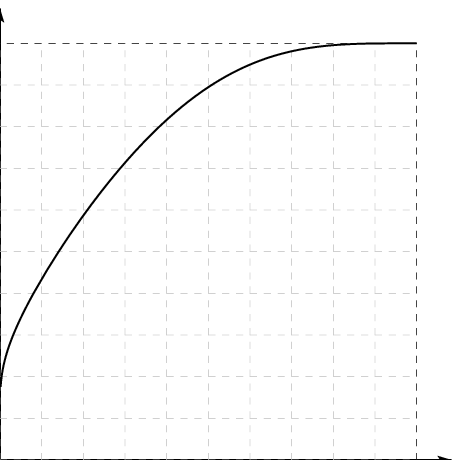}}
\put(60, -2){\makebox(0,0)[t]{\small $\entropy(\Ldens{c}_{\ent})$}}
\put(-2, 60){\makebox(0,0)[r]{\small \rotatebox{90}{$\gentropy(\Ldens{c}_{\ent}, \Ldens{a}_{\ent})$}}}
\put(60, 30){\makebox(0,0)[t]{\small standard GEXIT}}
\put(-2, -2){\makebox(0,0)[rt]{\small $0$}}
\put(120,-2){\makebox(0,0)[t]{\small $1$}}
\put(-2,120){\makebox(0,0)[r]{\small $1$}}
}
\put(140,0)
{
\put(0,0){\includegraphics[scale=1.0]{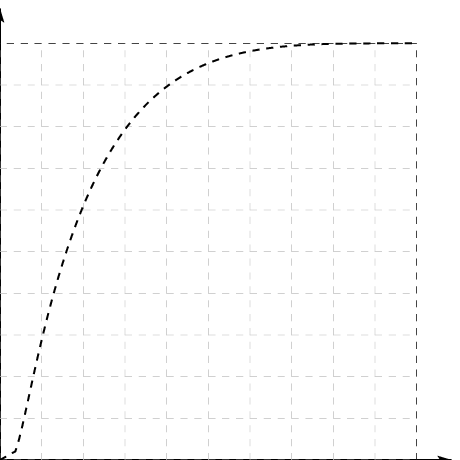}}
\put(60, -2){\makebox(0,0)[t]{\small $\gentropy(\Ldens{a}_{\ent}, \Ldens{c}_{\ent})$}}
\put(-2, 60){\makebox(0,0)[r]{\small \rotatebox{90}{$\entropy(\Ldens{a}_{\ent})$}}}
\put(60, 30){\makebox(0,0)[t]{\small dual GEXIT}}
\put(-2, -2){\makebox(0,0)[rt]{\small $0$}}
\put(120,-2){\makebox(0,0)[t]{\small $1$}}
\put(-2,120){\makebox(0,0)[r]{\small $1$}}
}
\put(280,0)
{
\put(0,0){\includegraphics[scale=1.0]{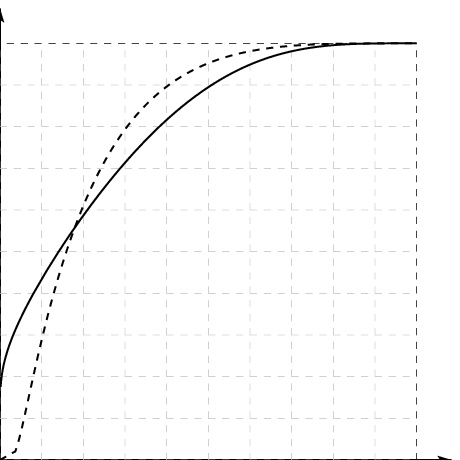}}
\put(60, -2){\makebox(0,0)[t]{\small $\entropy(\Ldens{c}_{\ent})$, $\gentropy(\Ldens{a}_{\ent}, \Ldens{c}_{\ent})$}}
\put(-2, 60){\makebox(0,0)[r]{\small \rotatebox{90}{$\gentropy(\Ldens{c}_{\ent}, \Ldens{a}_{\ent})$,$\entropy(\Ldens{a}_{\ent})$}}}
\put(60, 30){\makebox(0,0)[t]{\small both GEXIT}}
\put(-2, -2){\makebox(0,0)[rt]{\small $0$}}
\put(120,-2){\makebox(0,0)[t]{\small $1$}}
\put(-2,120){\makebox(0,0)[r]{\small $1$}}
}
\end{picture}
\end{center}
\caption{\label{fig:gexitanddualgexit} Standard and dual GEXIT function of $[5, 4, 2]$
code and transmission over the $\BSC$.}
\end{figure}
\end{example}
\blemma
Consider a degree distribution pair $(\ledge, \redge)$
and transmission over an BMS channel characterized by its
$L$-density $\Ldens{c}$ so that density evolution converges to 
$\Delta_{\infty}$. 
Let $\{\Ldens{a}_{\alpha}\}_{\alpha=-1}^{\infty}$
and $\{\Ldens{b}_{\alpha}\}_{\alpha=0}^{\infty}$ denote the interpolated
families as defined in Definition \ref{def:interpolation}.

Then the two GEXIT curves parameterized by
\begin{align*}
\{ \entropy(\Ldens{a}_{\alpha}), 
\gentropy(\Ldens{a}_{\alpha}, \Ldens{b}_{\alpha+1}) \}, \tag*{GEXIT of check nodes} \\
\{ \entropy(\Ldens{a}_{\alpha}),
\gentropy(\Ldens{a}_{\alpha}, \Ldens{b}_{\alpha}) \}, \tag*{inverse of dual GEXIT of variable nodes}
\end{align*}
do not overlap and faithfully represent density evolution.
Further, the area under the ``check-node'' GEXIT function
is equal to $1-\int \!\redge$ and the area to the left of the
``inverse dual variable node'' GEXIT function is equal to $\entropy(\Ldens{c}) \int \!\ledge$.
It follows that $r(\ledge, \redge) \leq 1-\entropy(\Ldens{c})$, i.e., 
the transmission rate can not exceed the Shannon limit.

This implies that transmission approaching capacity requires
a perfect matching of the two curves.
\elemma
\begin{proof}
First note that $\{ \entropy(\Ldens{a}_{\alpha}),
\gentropy(\Ldens{a}_{\alpha}, \Ldens{b}_{\alpha+1}) \}$
is the standard GEXIT curve representing the action
of the check nodes: $\Ldens{a}_{\alpha}$ corresponds to
the density of the messages {\em entering} the check nodes and
$\Ldens{b}_{\alpha+1}$ represents the density of the corresponding 
output messages.
On the other hand,
$\{ \entropy(\Ldens{a}_{\alpha}),
\gentropy(\Ldens{a}_{\alpha}, \Ldens{b}_{\alpha}) \}$
is the inverse of the dual GEXIT curve 
corresponding to the action at the variable nodes:
now the input density to the check nodes is 
$\Ldens{b}_{\alpha}$ and $\Ldens{a}_{\alpha}$ denotes the 
corresponding output density.

The fact that the two curves do not overlap can be seen as follows.
Fix an entropy value. This entropy value corresponds to a
density $\Ldens{a}_{\alpha}$ for a unique value of $\alpha$.
The fact that
$G(\Ldens{a}_{\alpha}, \Ldens{b}_{\alpha}) \geq
G(\Ldens{a}_{\alpha}, \Ldens{b}_{\alpha+1})$ now follows from
the fact that $\Ldens{b}_{\alpha+1} \prec \Ldens{b}_{\alpha}$ and
that for any symmetric $\Ldens{a}_{\alpha}$ this relationship
stays preserved by applying the GEXIT functional.

The statements regarding the areas of the two curves 
follow in a straightforward manner from the GAT and Lemma \ref{lem:dualgexit}.
The bound on the achievable rate follows in the same manner as for
the BEC: the total area of the GEXIT box equals one and the two curves do not
overlap and have areas $1-\int \redge$ and $\entropy(\Ldens{c})$.
It follows that
$1-\int \!\redge + \entropy(\Ldens{c}) \int \!\ledge \leq 1$,
which is equivalent to the claim $r(\ledge, \redge) \leq 1-\entropy(\Ldens{c})$.
\end{proof}

\begin{figure}[hbt]
\centering
\setlength{\unitlength}{1.5bp}
\begin{picture}(220,220)
\put(0,120){
\put(0,0){\includegraphics[scale=1.5]{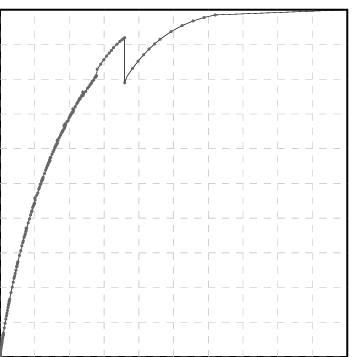}}
\put(50, -2){\makebox(0,0)[t]{\small $\entropy(\Ldens{a}_{\alpha})$}}
\put(102, 50){\makebox(0,0)[l]{\small \rotatebox{90}{$\gentropy(\Ldens{a}_{\alpha}, \Ldens{b}_{\alpha+1})$}}}
\put(50, 40){\makebox(0,0)[t]{\small GEXIT: check nodes}}
\put(50, 30){\makebox(0,0)[t]{\small $\text{area}=\frac56$}}
\put(50, 10){\makebox(0,0)[c]{$\Ldens{b}_{\alpha+1} = \sum_{i} \redge_i \Ldens{a}_{\alpha}^{\boxast (i-1)} $}}
}
\put(120, 120)
{
\put(0,0){\includegraphics[scale=1.5]{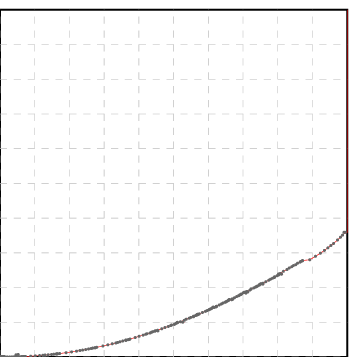}}
\put(50, -2){\makebox(0,0)[t]{\small {$\gentropy(\Ldens{a}_{\alpha}, \Ldens{b}_{\alpha})$}}}
\put(-2, 50){\makebox(0,0)[r]{\small  \rotatebox{90}{$\entropy(\Ldens{a}_{\alpha})$}}}
\put(50, 70){\makebox(0,0)[t]{\small dual GEXIT: variable nodes}}
\put(50, 60){\makebox(0,0)[t]{\small $\text{area}=\frac13 h(0.07)$}}
\put(102, 36.6){\makebox(0,0)[l]{\small \rotatebox{90}{$h(0.07) \approx 0.366$}}}
\put(50, 40){\makebox(0,0)[c]{$\Ldens{a}_{\alpha} = \Ldens{c} \conv \sum_{i} \ledge_i \Ldens{b}_{\alpha}^{\conv (i-1)} $}}
}
\put(0,0)
{
\put(0,0){\includegraphics[scale=1.5]{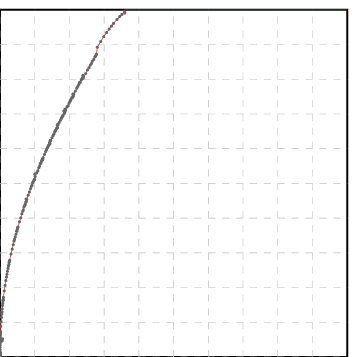}}
\put(50, -2){\makebox(0,0)[t]{\small $\entropy(\Ldens{a}_{\alpha})$}}
\put(-2, 50){\makebox(0,0)[r]{\small \rotatebox{90}{$\gentropy(\Ldens{a}_{\alpha}, \Ldens{b}_{\alpha})$}}}
\put(50, 30){\makebox(0,0)[t]{\small inverse of dual GEXIT:}}
\put(50, 20){\makebox(0,0)[t]{\small variable nodes}}
\put(36.6, 102){\makebox(0,0)[b]{\small $h(0.07) \approx 0.366$}}
}
\put(120,0)
{
\put(0,0){\includegraphics[scale=1.5]{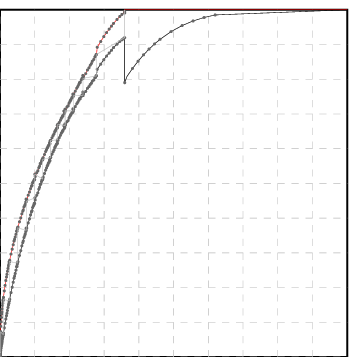}}
\put(50, -2){\makebox(0,0)[t]{\small $\entropy(\Ldens{a}_{\alpha})$}}
\put(-2, 50){\makebox(0,0)[r]{\small \rotatebox{90}{$\gentropy(\Ldens{a}_{\alpha}, \Ldens{b}_{\alpha})$}}}
\put(102, 50){\makebox(0,0)[l]{\small \rotatebox{90}{$\gentropy(\Ldens{a}_{\alpha}, \Ldens{b}_{\alpha+1})$}}}
\put(36.6, 102){\makebox(0,0)[b]{\small $h(0.07) \approx 0.366$}}
}
\end{picture}
\caption{
\label{fig:componentgexit}
Faithful representation of density evolution by two non-overlapping component-wise
GEXIT functions which represent the ``actions'' of the check nodes and variable nodes,
respectively. The area between the two curves equals is equal to the additive
gap to capacity.
}
\end{figure}

We see that the matching condition still holds even for general channels.
There are a few important differences between the general case and the simple
case of transmission over the BEC. For the BEC, the intermediate densities
are always the BEC densities independent of the degree distribution.
This of course enormously simplifies the task. Further, for the BEC, given
the two EXIT curves, the progress of density evolution is simply given
by a staircase function bounded by the two EXIT curves. For the general case,
this staircase function still has vertical pieces but the ``horizontal''
pieces are in general at an angle. This is true since the $y$-axis for
the ``check node'' step measures 
$\gentropy(\Ldens{a}_{\alpha}, \Ldens{b}_{\alpha+1})$, but
in the subsequent ``inverse variable node'' step 
it measures 
$\gentropy(\Ldens{a}_{\alpha+1}, \Ldens{b}_{\alpha+1})$.
Therefore, one should think of two sets of labels on the $y$-axis,
one measuring $\gentropy(\Ldens{a}_{\alpha}, \Ldens{b}_{\alpha+1})$,
and the second one measuring $\gentropy(\Ldens{a}_{\alpha+1}, \Ldens{b}_{\alpha+1})$. The ``horizontal'' step then consists of first
switching from the first $y$-axis to the second, so that the labels
correspond to the same density $\Ldens{b}$ and then drawing a horizontal
line until it crosses the ``inverse variable node'' GEXIT curve.
The ``vertical'' step stays as before, i.e., it really corresponds to
drawing a vertical line. All this is certainly best clarified by
a simple example.
\bex[$(3, 6)$-Regular Ensemble and Transmission over $\BSC$]
Consider the $(3, 6)$-regular ensemble and transmission over the $\BSC(0.07)$.
The corresponding illustrations are shown in Fig.~\ref{fig:componentgexit}.
The top-left figure shows the standard GEXIT curve for the check node side.
The top-right figure shows the dual GEXIT curve corresponding to the
variable node side. In order to use these two curves in the same figure,
it is convenient to consider the inverse function for the variable
node side. This is shown in the bottom-left figure. In the bottom-right
figure both curves are shown together with the ``staircase'' like function
which represents density evolution. As we see, the two curves to not overlap
and have both the correct areas.
\eex

As remarked earlier, one potential use of the matching condition
is to find capacity approaching degree distribution pairs. Let us
quickly outline a further such potential application. Assuming that
we have found a sequence of capacity-achieving degree distributions,
how does the number of required iterations scale as we approach capacity.
It has been conjectured that the the number of required iterations
scales like $1/\delta$, where $\delta$ is the gap to capacity.
This conjecture is based on the geometric picture which the
matching condition implies. To make things simple, imagine
the two GEXIT curves as two parallel lines, lets say both
at a 45 degree angle, a certain distance
apart, and think of density evolution as a staircase function.
From the previous results, the area between the lines is proportional
to $\delta$. Therefore, if we half $\delta$ the distance between
the lines has to be halved and one would expect that we need
twice as many steps. Obviously, the above discussion was 
based on a number of simplifying assumptions. It remains to 
be seen if this conjecture can be proven rigorously.

\section*{Acknowledgments}

The work of A.~Montanari was partially supported by the European Union under 
the project EVERGROW.
\bibliographystyle{ieeetr} 

\newcommand{\SortNoop}[1]{}

\end{document}